\def\year{2022}\relax
\documentclass[letterpaper]{article} 
\usepackage{aaai22}  
\usepackage{times}  
\usepackage{helvet}  
\usepackage{courier}  
\usepackage[hyphens]{url}  
\usepackage{graphicx} 
\usepackage{natbib}  
\usepackage{caption} 
\DeclareCaptionStyle{ruled}{labelfont=normalfont,labelsep=colon,strut=off} 
\usepackage{algorithm}
\usepackage{algorithmic}
\usepackage{tikz, pgfplots}
\usepackage{subcaption}
\usepackage{amsmath}
\usepackage{newfloat}
\usepackage{listings}
\floatstyle{ruled}
\newfloat{listing}{tb}{lst}{}
\floatname{listing}{Listing}
\title{JDRec: Practical Actor-Critic Framework for Online Combinatorial Recommender System }
\author {
    Xin Zhao,\textsuperscript{\rm 1}
    Zhiwei Fang, \textsuperscript{\rm 1}
    Yuchen Guo, \textsuperscript{\rm 2}
    Jie He, \textsuperscript{\rm 1}
    Wenlong Chen, \textsuperscript{\rm 1}
    Changping Peng \textsuperscript{\rm 1}
}
\affiliations {
    \textsuperscript{\rm 1} JD.com\\
    \textsuperscript{\rm 2} Tsinghua University\\
    zhaoxin19@gmail.com, fangzhiwei2@jd.com, yuchen.w.guo@gmail.com, hejie67@jd.com, chenwenlong17@jd.com, pengchangping@jd.com
}

\usepackage{bibentry}

\begin{document}

\maketitle

\begin{abstract}
A combinatorial recommender (CR) system feeds a list of items to a user at a time in the result page, in which the user behavior is affected by both contextual information and items. The CR is formulated as a combinatorial optimization problem with the objective of maximizing the recommendation reward of the whole list. Despite its importance, it is still a challenge to build a practical CR system, due to the efficiency, dynamics, personalization requirement in online environment. In particular, we tear the problem into two sub-problems, list generation and list evaluation. Novel and practical model architectures are designed for these sub-problems aiming at jointly optimizing effectiveness and efficiency. In order to adapt to online case, a bootstrap algorithm forming an actor-critic reinforcement framework is given to explore better recommendation mode in long-term user interaction. Offline and online experiment results demonstrate the efficacy of proposed JDRec framework. JDRec has been applied in online JD recommendation, improving click through rate by 2.6\% and synthetical value for the platform by 5.03\%. We 
will publish the large-scale dataset used in this study to contribute to the research community.
\end{abstract}

\section{Introduction}
Recommender System (RS) has attracted extensive research attention as long as the booming information growth on Mobile Internet. It shows significant importance on E-commerce scene for mining user interest. Recommender System is a case of Retrieval System without clear query input and homogeneous output. It takes user history behavior as query input and return diverse output to hit user interest. Early Recommender System takes basic methodology on ranking-based point-wise item recommendation \cite{GuoTYLH17,ZhouZSFZMYJLG18}. 

Point-wise item recommendation methods try to learn a correlation function between user and item with Collaborative Filter algorithm \cite{SarwarKKR01,KorenBV09,HeZKC16} or apply a key event probability prediction (e.g. CTR estimation \cite{GuoTYLH17,Cheng0HSCAACCIA16,WangFFW17,ZhouZSFZMYJLG18}) according to mixed features of user behavior and item attributes. Then the Recommender System operates according to the ranking list referred to correlation or probability. Point-wise item recommendation methods focus on the data-based metrics like AUC and NDCG, which indicate the ranking capability of point-wise model. Ranking-based point-wise item recommendation methods take a scoring and ranking pipeline, which is easier to apply on the online Recommender System with large amount of items.  The implicit assumption of point-wise item recommendation methods is that the combinatorial mode makes no difference on the chance of user click, and good item for a user is always good whatever the surrounding items are \cite{Cheng0HSCAACCIA16,WangFFW17}. They ignore the influence of other exposed items on the click through rate of anchor item, which limit the precision of estimation.

\begin{figure}[t]
  \centering
  \includegraphics[width=1\linewidth]{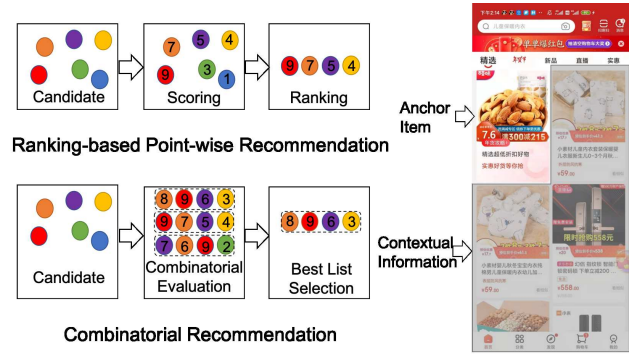}
  \caption{Point-wise vs Combinatorial Recommendation}\vspace{-0.6cm}
  
  \label{Combinatorial Recommendation}
  
\end{figure}

To improve the effectiveness of ranking-based point-wise recommendation methods, more and more attention are paid to combinatorial recommendation methods. Figure \ref{Combinatorial Recommendation} shows the difference between point-wise recommendation and combinatorial recommendation methods. Combinatorial recommendation evaluate candidate list as a whole in the mass and select the best list to expose. Recent research work shows that list diversity and intra-list correlations have great influence on user behavior. To improve the diversity of the list, a series of research works, e.g., MMR, IA-Select, xQuAD and DUM \cite{AgrawalGHI09,AshkanKBW15,CarbonellG98,SantosMO10}, are proposed to rank items by weighted functions to make a trade-off between user-item scores and diversities of items. However, these methods fail to take the impact of diversity on the list efficiency into consideration. A deep determinantal point process (Deep DPP) \cite{WilhelmRBJCG18} model 
is proposed to unify the user-item scores and the distances of items, which is a practical personalized approach. Yet simple optimization on list diversity is far from enough. Diversity is related to user experience but other user experience factors (e.g. category importance and correlation) is not taken into consideration. Diversity-based list optimization methods take long term user interest exploration into consideration but fail to find efficient exploration approach leading to a waste of exploration chance.

Experimental investigations show that contextual information (surrounding items) plays a significant role in prediction of user click action. Intra-item features are involved into ranking models to refine ranking scores in a n-element list \cite{AiBGC18,AiWBGBN19,ZhuangOW18}, which is commonly called "Slate Reranking". 
The Slate Reranking tries to present combinatorial recommendation from a global perspective instead of point-wise value estimation and sorting along with rule-based diversity control. That is to say, we need not put best at first in Slate Reranking, which is a popular way in the point-wise industrial online Recommender System. The optimized mode of list organization should be studied. There are two main challenges in Slate Reranking problem: a) how to generate candidate lists with better recommendation result; b) how to evaluate candidate lists and select the best one.

The procedures of reinforcement learning can be summarized as the following four steps: try, evaluate and feedback, and then update to do better, which is very suitable for exploring better recommendation mode. Generator and Critic framework is proposed \cite{abs-1902-00245,abs-2005-12206} in industrial Recommender System to solve the Slate Re-ranking problem, which belongs in Actor-Critic reinforcement method. These works give similar framework design and try to study the model structure on the list generator and list evaluator. RNN-based generator model and Attention-based evaluator model are proposed \cite{abs-1902-00245,abs-2005-12206}, which convert combinatorial recommendation problem into a sequential recommendation problem. It needs to be noted that the proposed Generator and Critic learning framework arises the following four problems:

\begin{itemize}
\item High time cost of RNN-based generator with multi-round neural network request.
\item Prediction bias on list evaluator when we change the recommendation mode from ranking-based one.
\item Initialization of list generator before push it online.
\item Controlling the exploration of list generator in the confidence region of list evaluator.
\end{itemize}

The above problems make it very difficult to transform the point-wise ranking-based recommendation framework to a Generator and Critic framework. In this work, we propose JDRec framework with novel model structure for List Generator to reduce time cost online, for which we take a bootstrap approach to solve the initialization problem. The core contributions in this work are are summarized as follows:

\begin{itemize}
\item We design a novel Set to List list generator along with list generation algorithm, which need just one neural network prediction for candidate list sampling. Compared to existing RNN-based list generation method it is so fast that we can use it online. 
\item We present a particular training method for the Set to List generator model to balance generation mode exploration and prediction result confidence. 
\item We propose a bootstrap approach which contains two steps to transform a ranking-based Recommender System to a reinforcement combinatorial one smoothly while we get better recommendation result in the whole procedure improving click through rate by 2.6\% and synthetical value for the platform by 5.03\%.
\item We publish a large-scale dataset from decision log of JD Recommender System reranking module, so that other researchers can do further research.
\end{itemize}

\section{Related Work}

\subsection{Context-aware Recommendation}
Recent research work shows that list diversity and intra-list correlations have great influence on user behavior in Recommender System. There exist two main categories of context-aware recommendation methods, which are point-wise method with contextual information added in and list-wise optimization through Slate Re-ranking. Contextual point-wise methods include deep determinantal point process (Deep DPP) \cite{WilhelmRBJCG18} and deep listwise context model(DLCM) \cite{AiBGC18} which model local contextual information and generate item list with greedy algorithm. Slate Re-ranking methods \cite{PeiZZSLSWJGOP19,abs-1902-00245,abs-2005-12206} are proposed to make global list-wise optimization for better recommendation effect. Generator and Critic framework \cite{abs-1902-00245,abs-2005-12206} shows great potential on solving Slate Re-ranking problem with an reinforcement paradigm.

\begin{figure*}[t]
  \centering
  \includegraphics[width=0.7\linewidth]{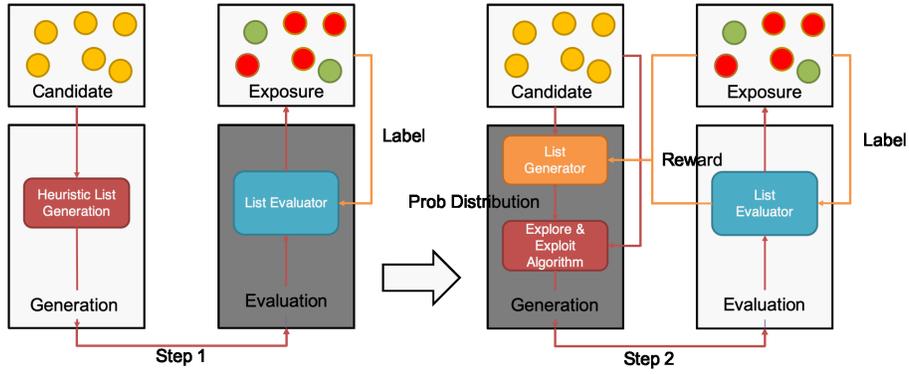}
  \caption{Two-step bootstrap method to run JDRec framework where red lines represent forward procedure and yellow lines represent backward procedure. The gray boxes represent core components in each step.}\vspace{-0.5cm}
  \label{Bootstrap}
\end{figure*}

In this work, we make great improvement on Generator and Critic framework and propose a novel model structure and two-step pipeline to make it serve online which balances online time cost and recommendation profit.
\subsection{Reinforcement Learning}
In recent years, Reinforcement Learning has shown great power on approximately solving complex Combinatorial Optimization Problem(e.g., Go, Atari Game). It outperforms human when the rule is simple or the feedback of environment can be acquired easily. Reinforcement learning methods take exploration and learn to make
better decision iteratively. Actor-Critic framework \cite{SilverLHDWR14} is proposed for single step policy update and faster reinforcement learning which makes it easier to solve decision problems on long action sequence. It is found that reinforcement learning model can only preform well in the border-upon domain of observed samples called trust region. Trust region learning methods (e.g., TRPO \cite{SchulmanLAJM15} and PPO \cite{SchulmanWDRK17}) are proposed to take more effective exploration and monotonic learning.

The Generator and Critic framework for context-aware recommendation is a case of Actor-Critic method. And in this work, we take the trust region exploration problem into consideration and take a controlled learning approach on list generator for online risk control.

\section{Methodology}
In this section, we present the proposed JDRec framework to solve the combinatorial recommendation problem. 
\subsection{Problem Formulation}
The combinatorial recommendation problem can be formulated as follows. Given a candidate item set $\mathcal{I}=\{I_1, I_2, I_3, …, I_n\}$ and user profile $\mathcal{U}$, the combinatorial recommendation problem is to find an optimized sequence $\{\mathcal{A}\} = \{I_{a_1}, I_{a_2}, …, I_{a_l}\}$, where $I_{a_i} \in \mathcal{I}$. Then, $\{\mathcal{A}\}$ is exposed to user, after which the user returns the feedback of $r(\mathcal{U},\mathcal{A})$. We define the final objective as maximizing the expected overall utility of the sequence $\mathcal{A}$, which is written as $E[r(\mathcal{U},\mathcal{A})]$, where $E_X[]$ represents the expectation over variable X, and $E[]$ represents the expectation over repeated experiments.
\begin{equation}
  \mathcal{A}_{opt} = argmax_\mathcal{A}(E[r(\mathcal{U},\mathcal{A})]).
\end{equation}
Multiple feedbacks are available in Recommender System except for clicks (e.g. order chance). The reward function $r(\mathcal{U},\mathcal{A})$ is a combination of all the feedback rewards.
\subsection{JDRec Framework}
In this section we introduce the proposed JDRec framework, which is a variant of Actor-Critic reinforcement learning framework for Recommender System. 

In forward procedure, we are given a candidate item set and it is fed into the List Generator. The List Generator predicts a policy probability distribution for list generation. Then, we take a list generation sampling algorithm for both exploring new action and exploiting reward, which generate candidate item lists for selection. Finally, we feed the generated candidate item lists into List Evaluator and select the best list for exposure according to its expectation on reward function $E[r(\mathcal{U},\mathcal{A})]$.

In backward procedure, we get the user behavior online and compute the reward value according to it. The user behavior provides labels to List Evaluator to improve the prediction accuracy. Online reward and the evaluation result provide a mixed reward for updating List Generator. 

The list generation algorithm is a Markov Chain Monte Carlo (MCMC) sampling method which is controlled by a temperature parameter. We will introduce this sampling algorithm afterwards.

Theoretically, if the JDRec framework can work on the online Recommender System, the List Evaluator can learn directly from online user behavior for more accurate prediction on the true action space. Then, it will give better reward estimation to the list generator for better generator policy, which is a reinforcement circle. However, as discussed before, the initialization for List Generator is difficult so that it is almost impossible for us to push both List Generator and List Evaluator online at the same time. Therefore we propose a novel bootstrap method to change the online ranking-based Recommender System step by step, which includes two main steps for smoothly changing the online system with better recommendation result in each step. The two-step method makes it possible to run a reinforcement loop on practical Recommender System as shown in Figure \ref{Bootstrap}. 

\begin{figure}[h]
  \centering
  \includegraphics[width=1\linewidth]{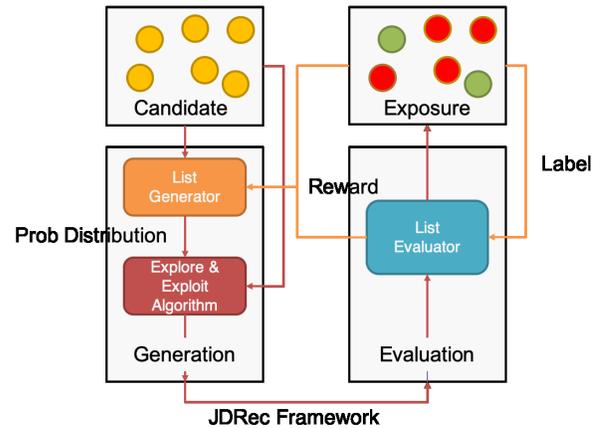}
  \caption{Ideal JDRec framework where red lines represent forward and yellow lines represent backward procedure.}\vspace{-0.2cm}
  \label{JDRec}
\end{figure}

The initialization of data-driven list generation model is very hard before we change the ranking-based recommendation strategy, because all the explorations are in a single policy space so the observation of list generator is biased. We have to change the online recommendation strategy first to train a model-based List Generator. And the recommendation mode should be diverse. At the same time, we cannot afford bad recommendation results online. So we design first step for JDRec framework, with a heuristic list generation algorithm and a model-based List Evaluator. Heuristic list generation algorithm is a combination of several generation methods including ranking-based list generation and random sampling methods according to point-wise ranking score, which guarantees the lower bound of the best list in the candidate set is no worse than original ranking-based recommendation method. Then, the candidate list set is fed into List Evaluator to select the best list for exposure. The List Evaluator is biased at the beginning, so we take Gray Release to de-bias on List Evaluator gradually.

After step 1 is finished, we get diversiform exploration on recommendation mode, and the accumulated data can be used to train a model-based List Generator. Specifically, we can train a List Generator with a candidate item set as input and list generation probability distribution as output. And we take a MCMC sampling algorithm for both exploration and exploitation. The model structures and sampling algorithm will be stated afterwards.

\subsection{List Evaluator}
\begin{figure}[h]
  \centering
  \includegraphics[width=0.8\linewidth]{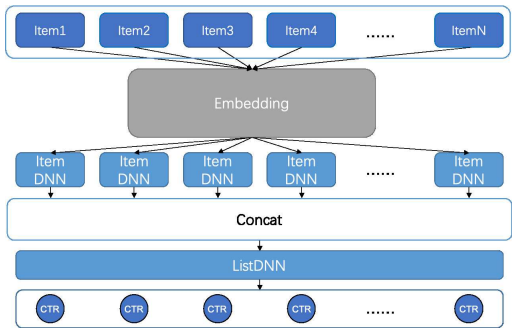}
  \caption{List Evaluator.}\vspace{-0.2cm}
  \label{Evaluator}
\end{figure}

The goal of the List Evaluator is to predict CTR for each item in the list. The model structure of the proposed List Evaluator is shown in Figure \ref{Evaluator}.

The output of the List Evaluator is the overall click rate of the list. The input of the List Evaluator is a sequence of items with user interactive features in a user session, point-wise prediction result and other side information of item. The List Evaluator converts the input features to a normal embedding vector with shared embedding parameters. Then, DNN layers are applied to the embedding vector to extract item representation. Next, a concat layer is applied to get a list feature vector while several DNN layers are applied to predict Click Through Rate(CTR) for each item.

\textbf{Training Details.} The training samples are collected from real online request log, while the list samples are not necessarily to expose to user because of follow-up filter logic and user interactive action. The training procedure of List Evaluator can be formulated as follow. Given an item list $\{\mathcal{A}\} = \{I_{a_1}, I_{a_2}, …, I_{a_l}\}$ and its exposure label $\{\mathcal{L}_{exposure}\} = \{L_{exposure, a_1}, L_{exposure, a_2}, …, L_{exposure, a_l} \}$ and click label $\{\mathcal{L}_{click}\} = \{L_{click, a_1}, L_{click, a_2}, …, L_{click, a_l}\}$, where $L_{exposure, a_i} \in \{0, 1\}$ and $L_{click, a_i} \in \{0, 1\}$, where $1$ means exposure and click respectively. The aim of List Evaluator training is to estimate CTR on the candidate lists. Because the unexposed items do not have chance to be click, we ignore the loss computation on them by using exposure label. The loss function of List Evaluator is as follows.
\begin{equation}
\begin{aligned}
  Loss_{evaluator} = \sum_{i=1}^l L_{exposure, a_i}[L_{click, a_i}log(\hat{p}_{a_i}) \\
  + (1 - L_{click, a_i})log(1 - \hat{p}_{a_i})], 
\end{aligned}
\end{equation}
which is a weighted sigmoid cross entropy loss where $\hat{p}_{a_i}$ represents the Click Through Rate for item $I_{a_i}$ predicted by the List Evaluator. We take AdaGrad algorithm with initial learning rate 0.01 to train List Evaluator.

\subsection{List Generator}
\begin{figure}[t]
  \centering
  \includegraphics[width=0.8\linewidth]{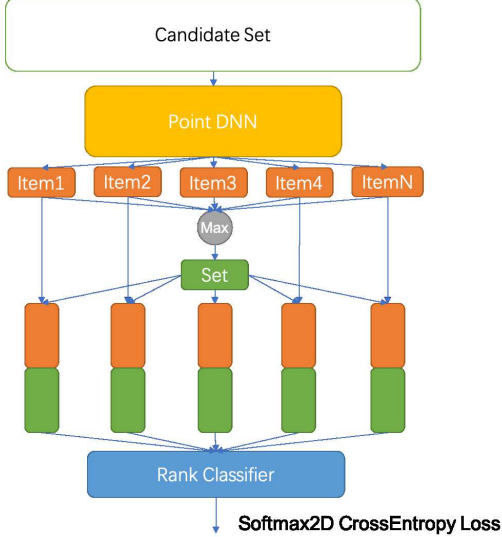}
  \caption{Set to Policy List Generator.}\vspace{-0.5cm}
  \label{Generator}
\end{figure}

The goal of the List Generator is to model a mapping from candidate set to list generation policy probability matrix. The model structure of the proposed Set to Policy List Generator is shown in Figure \ref{Generator}.

The existing Generator and Critic methods tend to model list generation procedure as a recurrent generation paradigm. They model list generation as a conditional random event, where the sampling probability for current item is determined by all the previous items in the list. The recurrent generation paradigm is an ideal way for list generation except the high time cost online. The proposed Set to Policy List Generator is an approximate approach for the real-time recommendation without multi-round model prediction.

The input of the List Generator is a candidate item set with features, which are related to the user interactive action and our optimization object (e.g., CTR, order chance and improve weight). The candidate set is permutation-invariant, so that the Set to Policy list generation model must be irrelevant to the order of items in candidate set. The aim of the proposed List Generator is to give the probability in the list generation procedure, so that we can take advantage of the distribution to sample several candidate item lists which seem to be good. The Set to Policy List Generator can be regarded as a selector from candidate set. 

The output of the Set to Policy List Generator is a $(L+1) \times N$ 2D matrix $M$ where N represents the size of candidate set and L represents the length of generated candidate list. Here is an example of the generator output in Table \ref{Generator Output}.
\begin{table}
\centering
  \begin{tabular}{c|cccccc}
    &1&2&3&4&5&6\\
    \hline
    1 & 0.9 & 0.1 & 0  & 0  & 0 & 0\\
    2 & 0.05& 0.8 &0.15& 0  & 0 & 0\\
    3 & 0.05& 0.1 & 0.7&0.15& 0 & 0\\
    4 & 0   &  0  &0.15&0.85& 0 & 0\\
    Not in & 0   &  0  & 0  & 0  & 1 & 1\\
\end{tabular}
\caption{Example of List Generator Output with L=4, N=6}\vspace{-0.5cm}
\label{Generator Output}
\end{table}

In specific, $M[i,:]$ represents the probability for each candidate item to be in position $i$, where $i \in \{1,… ,L\}$. $M[:,j]$ represents the probability for j-th item to be in for each position in the list, including not in the list. Once we get the matrix, we can sample item in the candidate set one by one according to $M[i,:]$. The details of the list generation algorithm will be presented afterwards.

The Set to Policy List Generator first uses a Point DNN sub-graph to draw feature vector for each item, whose structure is similar to the List Evaluator. And then it applies a max-pooling layer to fuse them as the feature of the whole candidate set. Next, the feature of candidate set is concatenated to the item features. The concatenation of set feature and item feature contains necessary information for item rank in candidate set. Finally, a Rank Classifier is built to predict the probability of item in each position as Table \ref{Generator Output}.

\textbf{Training Details.} The label for the Set to Policy List Generator is ordered id list in the candidate list selected by the List Evaluator, whose sparse representation is $IDs$ = $\{id_1, id_2, …, id_L\}$.  And $IDs$ can be easily converted to rank label $Rank(x)$ for each item as follows.

\begin{equation}
Rank(x) = 
\begin{cases}
j &x=IDs[j] \\
L+1 & others
\end{cases}.
\end{equation}

The proposed Softmax2D cross entropy loss is actually a mixture of two sub-task, which are id selection for each position and rank classification for each candidate item. And the loss function for these two sub-task are denoted by $Loss_{id}$ and $Loss_{rank}$, which are formulated as:

\begin{equation}
  Loss_{id} = \sum_{i=1}^L SoftmaxCrossEntropy(M[i,:], IDs[i]),
\end{equation}

\begin{equation}
  Loss_{rank} = \sum_{j=1}^N SoftmaxCrossEntropy(M[:,j], Rank(j)),
\end{equation}
where SoftmaxCrossEntropy(x,y) represents the common single-class classification loss, and the notation : represents the whole row or column in a particular position.

And the proposed Softmax2D cross entropy loss is a linear combination of them:
\begin{equation}
  Loss_{Softmax2D} = Loss_{id} + \lambda Loss_{rank}.
\end{equation}

\subsection{List Generation Algorithm}
In this section, we introduce the list generation algorithm to generate candidate list. We take a MCMC sampling algorithm with a temperature parameter $t$ to balance exploration and exploitation in list generation. The input of the algorithm is the 2D table $M$ from List Generator with shape $(L+1) \times N$, temperature $t$, list length $L$ and the number of max sample times $k$. We apply a multi-round list generation algorithm according to the following function:
\begin{equation}
  prob[i,j] = \frac{e^{tM[i,j]}}{\sum_{m=1}^N e^{tM[i,m]}}, \ i \leq L.
  \label{prob}
\end{equation}

MCMC generation algorithm is shown as Algorithm \ref{MCMC}.

\begin{algorithm}[tb]
\caption{MCMC List Generation Algorithm} 
\label{MCMC}
\hspace*{0.02in} {\bf Input:} 
input parameters M, t, k, L\\
\hspace*{0.02in} {\bf Output:} 
candidate list set s
\begin{algorithmic}[1]
\STATE set m = 1
\STATE set s = $\phi$
\FOR{m $\leq$ k}
    \STATE calculate table prob as function \ref{prob}
    \STATE set list = $\phi$
    \STATE set n = 1
    \FOR{n $\leq$ L}
        \WHILE{True}
            \STATE sample l according to prob[n,:]
            \IF{CheckLegal(list, l)}
                \STATE break
            \ELSE
                \STATE set prob[n,l] = 0
            \ENDIF
        \ENDWHILE
        \STATE add l to list
        \STATE set prob[:,l] = 0
        \STATE re-normalize prob from row
        \STATE set n = n + 1
    \ENDFOR
    \IF{list $\notin$ s}
        \STATE add list to s
    \ENDIF
    \STATE set m = m + 1
\ENDFOR
\RETURN s
\end{algorithmic}
\end{algorithm}

\section{Experimental Results}

\subsection{JDRec Dataset}
Our experiment dataset, named JDRec dataset, is sampled from the decision log of JD Recommender System, which is one of the largest e-commerce Recommender System in the world. JDRec dataset is the first large-scale combinatorial item recommendation dataset. JDRec dataset contains more than \textbf{7.52 million samples}. The whole JDRec dataset is divided into two parts which are train set and test set. The number of samples in these parts are 7.37 million and 153 thousand respectively. Each sample includes 40 candidate items and the 4 items finally selected by the online rerank module. Each item info consists of 2 labels and 51 features. The labels include  \textit{click} and \textit{RerankIndex}, which means whether the item is clicked by the user or not and the rerank position for the selected items with a default value for other candidate items. In order to avoid the disclosure of sensitive information, we have taken the following information protection procedures:
\begin{itemize}
    \item The JDRec dataset is sampled from a much more larger dataset from JD Recommender System randomly.
    \item Sensitive category features are being shuffled and renumbered.
    \item Sensitive numeric features (e.g. pctr) are transformed by a linear equation.
    \item Other features keep their origin value in the system log.
\end{itemize}

\subsection{Offline Evaluation}
In this section, we introduce the offline evaluation on the JDRec Dataset, including offline metrics of List Evaluator and List Generator.

\textbf{Metircs.} We take \textbf{AUC(Area Under Curve)} on exposed items as the evaluation metric for List Evaluator. To evaluate List Generator, it is hard to simulate the real online recommendation environment. So we train an offline List Evaluator as an interactive simulation environment for List Generator. And we simply take average click through rate on generated item lists for optimization. The evaluation metric for List Generator is \textbf{Average CTR} predicted by a fixed offline List Evaluator. 

\textbf{Experimental Settings.} The online ranking-base Recommender System of JD relies on a point-wise CTR estimation model which is a Deep and Cross Network (DCN) \cite{WangFFW17} with Transformer \cite{VaswaniSPUJGKP17} to model user history behavior. The CTR value from the point-wise model is feed into the List Evaluator to get more accurate re-estimation. The online point-wise CTR estimation model is taken as our baseline. List Evaluator and List Generator share the same feature column setting: All categorical features are being indexed and represented by an embedding matrix. Embedding dimensions of different features are various according to the value space of each feature. Every numerical features are being bucketized and represented by a embedding matrix. Several origin value of the numerical features are also used in the models for better training effect.

We take two kinds of training methods for List Generator termed as \textbf{Naive RL} and \textbf{CTR RL}. The difference between these two methods are on sampling algorithm and reward function. \textbf{Naive RL} only sample the item list selected in the online Recommender System and the reward function is 1 if the list is selected by online Recommender System and 0 for otherwise. The reward function of \textbf{CTR RL} is average predicted CTR of the list.

We sample item list according to the output matrix of List Generator and take policy gradient algorithm for reinforcement update. And we additionally sample the online list and the top ctr list for training acceleration. The experiment code for the offline evaluation is on \textbf{https://github.com/SeekerYb/JDRec}, where we public a script to download the JDRec dataset.
\begin{table}[ht]

\centering
  \begin{tabular}{lc}
    &AUC\\
    \hline
    point-wise CTR model & 0.7201\\
    List Evaluator  & 0.7248\\
    Diff & 0.0047\\

\end{tabular}
\caption{Evaluation on List Evaluator}\vspace{-0.3cm}
\label{OfflineEvaluator}
\end{table}

\textbf{Evaluation Result.} The evaluation results of List Evaluator and point-wise baseline model are listed in Tab \ref{OfflineEvaluator}. We obtain that the list-wise model brings \textbf{0.0047} growth on AUC compared with online point-wise CTR estimation model, which shows the advantage of List Evaluator over point-wise CTR estimation model.

The evaluation results of List Generator are listed in Tab \ref{OfflineGenerator}. We obtain that the CTR RL method makes remarkable improvement on click through rate of recommendation list when we take CTR-oriented reward function, which can be a baseline for further research on reinforcement learning for Recommender System. The experiment result shows that when we take CTR as the only aim we want to optimize by reinforcement learning, we can get much better result on this single index. 
\begin{table}[h]
\centering
  \begin{tabular}{lc}
    &Average CTR\\
    \hline
    Naive RL & 0.0390\\
    CTR RL & 0.0650\\
    Gain & 0.0260\\
    Relative Gain & 66.67\%\\
\end{tabular}
  \caption{Evaluation on List Generator}\vspace{-0.5cm}
  \label{OfflineGenerator}

\end{table}

\subsection{Online Evaluation}
In this section, we introduce the online evaluation of our JDRec Framework, including evaluation on List Evaluator and List Generator and A/B test results for 2 main steps. Besides, we report several long term monitor index to observe the reinforcement procedure.

\textbf{Metrics.} Same as offline evaluation, we take \textbf{AUC(Area Under Curve)} on exposed items as the evaluation metric for List Evaluator. Training of List Generator can be actually divided into 2 sub-tasks, which are item selection task on given position of list and rank prediction task for each candidate item. We take classification accuracy for these two sub-tasks as the evaluation metric for List Generator. The \textbf{Item Selection Accuracy} can be calculated from rows of $M$ while the \textbf{Rank Accuracy} can be calculated from columns of $M$. In the practical online Recommender System, to keep robustness of list generation we run model-based list generation and heuristic list generation at the same time. And then we select best list from the generated candidate list set. We use \textbf{Winning Rate} for a generation method to represent the relative metric from others. Higher \textbf{Winning Rate} means more best lists are generated by this method, which implies better list generation result than others.

\textbf{Experimental Settings.}  We will report AUC of the online point-wise model and the proposed List Evaluator to show the gain of ranking ability. The dates we selected are 21st May 2020 and 9th Dec 2020 which are the gray release dates for List Evaluator and List Generator. The model is our serving model online and the test set is from the practical logs of JD Recommender System, which is different from offline dataset. The online List Evaluator and List Generator will reload previous model and retrain day by day with current training log. Winning Rate is not a direct evaluation for the whole reinforcement system, but we cannot stop other optimization in the online Recommender System so that the direct online evaluation (e.g. total click) cannot show the effect of reinforcement learning fairly. We fix the heuristic list generation algorithm so that the change of Winning Rate can show the quality of list generation during the reinforcement learning procedure of List Generator. The online A/B test result is from online experiment during online algorithm hold-back period when 95\% of online requests are for experiment and 5\% for baseline.
\begin{figure}
    \centering
  \begin{tikzpicture}[scale = .6]
\begin{axis}[
    xlabel={Day},
    ylabel={Accuracy},
    xmin=0, xmax=30,
    ymin=0.4, ymax=0.6,
    xtick={0,5,10,15,20,25,30},
    ytick={0.4,0.45,0.5,0.55,0.6},
    legend pos=north west,
    ymajorgrids=true,
    grid style=dashed,
]
 
\addplot[
    color=blue,
    mark=square,
    ]
    coordinates {
    (0,0.436)(1,0.455)(2,0.452)(3,0.466)(4,0.458)(5,0.465)(6,0.469)(7,0.474)(8,0.47)(9,0.467)(10,0.479)(11,0.475)(12,0.479)(13,0.493)(14,0.487)(15,0.472)(16,0.476)(17,0.467)(18,0.462)(19,0.462)(20,0.538)(21,0.548)(22,0.568)(23,0.574)(24,0.581)(25,0.573)(26,0.566)(27,0.579)(28,0.596)(29,0.592)
    };
    \legend{Item Selection Accuracy}
 
\end{axis}
\end{tikzpicture}
\caption{Accuracy to Predict Item Given Position}\vspace{-0.5cm}
\label{ItemSelectionAccuracy}
\end{figure}
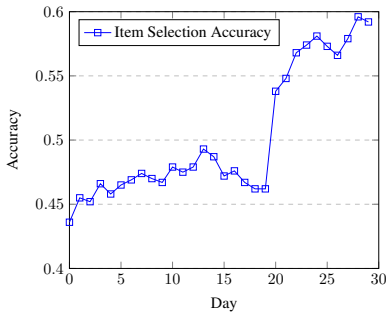

\textbf{Evaluation Result.} The evaluation results of List Evaluator and point-wise baseline model are listed in Tab \ref{AucOnline}. When the List Evaluator start to gray release the gain in AUC is \textbf{0.0084}. And then the number shrink to \textbf{0.0051} because of the optimization of baseline model when List Generator start to gray release. The A/B test result is shown in Tab \ref{ABTest}, where total value means synthetical value brought by user click (e.g. order and income). As is shown in Tab \ref{ABTest}, the JDRec framework brings immediate gain for online recommendation effect to make sure the procedure of releasing JDRec framework is smooth enough.

\begin{table}[h]
\centering
  \begin{tabular}{lcccccc}
    &21st May 2020&9th Dec 2020\\
    \hline
    point-wise CTR model & 0.7409 & 0.7181 \\
    List Evaluator       & 0.7493& 0.7232\\
    Diff                 & 0.0084& 0.0051\\

\end{tabular}
  \caption{AUCs of Online List Evaluator in Different Days}\vspace{-0.3cm}
  \label{AucOnline}
\end{table}

\begin{table}[h]
\centering
  \begin{tabular}{lcc}
    &click&total value\\
    \hline
    Evaluator (21st May) & +2.16\%& +3.68\% \\
    Generator (9th Dec)  & +0.44\%& +1.35\%\\
\end{tabular}
  \caption{A/B Test Results}\vspace{-0.3cm}
  \label{ABTest}
\end{table}

\begin{figure}
    \centering

  \begin{tikzpicture}[scale = .6]
\begin{axis}[
    xlabel={Day},
    ylabel={Accuracy},
    xmin=0, xmax=30,
    ymin=0.9, ymax=0.94,
    xtick={0,5,10,15,20,25,30},
    ytick={0.9,0.91,0.92,0.93,0.94},
    legend pos=north west,
    ymajorgrids=true,
    grid style=dashed,
]
 
\addplot[
    color=blue,
    mark=square,
    ]
    coordinates {
    (0,0.92)(1,0.923)(2,0.923)(3,0.924)(4,0.923)(5,0.924)(6,0.924)(7,0.924)(8,0.925)(9,0.924)(10,0.925)(11,0.924)(12,0.925)(13,0.926)(14,0.925)(15,0.924)(16,0.924)(17,0.923)(18,0.923)(19,0.923)(20,0.927)(21,0.928)(22,0.929)(23,0.929)(24,0.929)(25,0.929)(26,0.929)(27,0.93)(28,0.931)(29,0.93)
    };
    \legend{Rank Accuracy}
\end{axis}
\end{tikzpicture}
    \caption{Accuracy to Predict Rank Given Item}\vspace{-0.3cm}
    \label{RankAccuracy}
\end{figure}
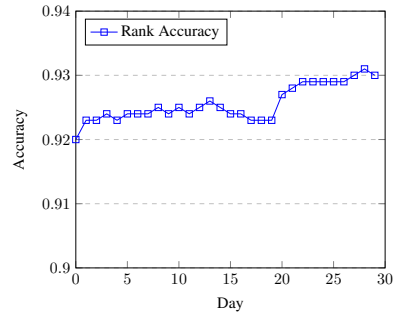

\begin{figure}
    \centering

\begin{tikzpicture}[scale = .6]
\begin{axis}[
    xlabel={Day},
    ylabel={Wining Rate},
    xmin=0, xmax=30,
    ymin=0.75, ymax=0.9,
    xtick={0,5,10,15,20,25,30},
    ytick={0.75,0.8,0.85,0.9},
    legend pos=north west,
    ymajorgrids=true,
    grid style=dashed,
]
 
\addplot[
    color=blue,
    mark=square,
    ]
    coordinates {
    (0,0.79)(1,0.791)(2,0.795)(3,0.793)(4,0.794)(5,0.793)(6,0.796)(7,0.797)(8,0.796)(9,0.798)(10,0.799)(11,0.8)(12,0.801)(13,0.803)(14,0.802)(15,0.805)(16,0.806)(17,0.807)(18,0.807)(19,0.806)(20,0.82)(21,0.825)(22,0.828)(23,0.83)(24,0.833)(25,0.835)(26,0.834)(27,0.838)(28,0.85)(29,0.845)
    };
    \legend{Winning Rate}
 
\end{axis}
\end{tikzpicture}
\caption{Winning Rate for List Generator}\vspace{-0.5cm}
\label{WinningRate}
\end{figure}
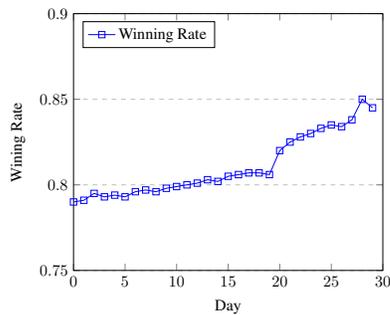

Figures \ref{ItemSelectionAccuracy} \ref{RankAccuracy} \ref{WinningRate} show the trend of online indexes around the gray release date for List Generator. The x-axis is for day. And the 20th day is 9th Dec 2020, when the request ratio for List Generator rise to \textbf{95\%} from \textbf{5\%}. In all the figures, we can see that there is a jump on the 20th day. And then the Item Selection Accuracy, the Rank Accuracy and the Winning Rate of model-based candidate list generation grows slowly. The online indexes show the reinforcement learning process of the proposed List Generator. That is to say, the ability for List Generator to guess what is a good list grows day by day and it is confirmed by the List Evaluator so that the model-based list generation method win more gradually. 

\section{Conclusion}
The key problem we have solved is how to apply Actor-Critic framework on the practical online Recommender System and achieve better performance. We design model structure for List Evaluator to take contextual information into consideration and improve the prediction accuracy for CTR estimation. The problem in Combinatorial Recommender System is a multi-step decision problem. We cannot afford a step-by-step decision process because of the real-time limitation of online system. So we convert it to a single-step distribution prediction problem and design a customized model structure for List Generator. And then we propose matched training algorithm and list generation algorithm for it. More importantly, we propose a bootstrap method to release the JDRec framework step by step, making it to run in the real industrial online system and achieve real reinforcement loop for it. We public part of the decision log of rerank module on JD Recommender System and provide a baseline result on it, which is the first large-scale dataset for combinatorial recommendation problem.

\bibliography{aaai22}

\end{document}